\documentclass[twocolumn,conference]{IEEEtran}
\usepackage[LGR,T1]{fontenc}
\usepackage{amsbsy}
\usepackage{graphicx}
\usepackage[unicode=true,
 bookmarks=true,bookmarksnumbered=true,bookmarksopen=true,bookmarksopenlevel=1,
 breaklinks=false,pdfborder={0 0 0},pdfborderstyle={},backref=false,colorlinks=false]
 {hyperref}
\hypersetup{pdftitle={Your Title},
 pdfauthor={Your Name},
 pdfpagelayout=OneColumn, pdfnewwindow=true, pdfstartview=XYZ, plainpages=false}

\makeatletter

\DeclareRobustCommand{\greektext}{%
  \fontencoding{LGR}\selectfont\def\encodingdefault{LGR}}
\DeclareRobustCommand{\textgreek}[1]{\leavevmode{\greektext #1}}
\ProvideTextCommand{\~}{LGR}[1]{\char126#1}

\usepackage[caption=false,font=footnotesize]{subfig}
\usepackage{amsmath}
\usepackage[linesnumbered,ruled]{algorithm2e}

\makeatother

\begin{document}

\title{Concatenated LDPC-Polar Codes Decoding Through Belief Propagation}

\author{\IEEEauthorblockN{Syed Mohsin Abbas, YouZhe Fan, Ji Chen and Chi-Ying Tsui}\IEEEauthorblockA{VLSI Research Laboratory, Department of Electronic and Computer Engineering\\
Hong Kong University of Science and Technology (HKUST), Hong Kong\\
Email: \{smabbas, jasonfan, jchenbh, eetsui\}@ust.hk}}
\maketitle
\begin{abstract}
Owing to their capacity-achieving performance and low encoding and
decoding complexity, polar codes have drawn much research interests
recently. Successive cancellation decoding (SCD) and belief propagation
decoding (BPD) are two common approaches for decoding polar codes.
SCD is sequential in nature while BPD can run in parallel. Thus BPD
is more attractive for low latency applications. However BPD has some
performance degradation at higher SNR when compared with SCD. Concatenating
LDPC with Polar codes is one popular approach to enhance the performance
of BPD , where a short LDPC code is used as an outer code and Polar
code is used as an inner code. In this work we propose a new way to
construct concatenated LDPC-Polar code, which not only outperforms
conventional BPD and existing concatenated LDPC-Polar code but also
shows a performance improvement of 0.5 dB at higher SNR regime when
compared with SCD. 
\end{abstract}


\section{Introduction}

Polar codes, since their invention by \emph{Ar$\imath$kan} \cite{arikan},
have been proven to achieve the capacity for binary-input symmetric
memory-less channels \cite{arikan} as well as discrete and continuous
memory-less channels \cite{arikan2}. Moreover, an explicit construction
method for polar codes has been provided, and it is shown that they
can be efficiently encoded and decoded with complexity $\mathcal{O}(n\,log\,n)$,
where $n$ is the code length. A number of decoding methods have been
proposed for polar codes \cite{yazdi}-\cite{Jun_lin}, and among
these successive cancellation decoding (SCD) and belief propagation
decoding (BPD) are the two most popular methods because of their performance
and easy implementation in hardware. 

Due to the serial nature of the algorithm, SCD suffers from longer
latency. BPD, on the other hand, has the intrinsic advantage of parallel
processing. Therefore, compared with SCD, BPD is more attractive for
low-latency applications. In \cite{TVLSI}, a high throughput BPD
(13.9Gbps) was proposed for (1024,512) polar codes which has an average
decoding latency of 37.8 cycles with a maximum frequency of 515MHz.
However despite its high throughput and lower latency, BPD suffers
from performance degradation at higher SNR when compared with SCD
\cite{TVLSI,Jun_lin}. To improve the performance of BPD, several
concatenated Polar codes with other block codes have been suggested
in literature. Eslami et al. \cite{Eslami} proposed to concatenate
polar codes and LDPC codes, both of long code lengths ($2^{15}$),
to be used in Optical Transport Network (OTN). This concatenated polar-LDPC
code has been shown to outperform LDPC code of the same length at
the cost of higher decoding complexity.

For smaller complexity overhead, Guo et al. \cite{Guo} proposed to
employ a short LDPC code as an outer code and larger Polar code as
an inner code in the concatenated code. This concatenated LDPC-polar
code results in 0.3dB improvement over standard BP decoding of polar
code. On the similar note, for smaller complexity overhead, in this
work we propose an alternate way to concatenate a short LDPC outer
code with a larger inner polar code. Our proposed concatenated LDPC-polar
code not only outperforms SCD and conventional BPD, but also achieves
performance improvement of 0.25dB and 0.1dB, at higher SNR regime,
when compared with existing concatenated LDPC-polar code \cite{Guo}
and list SCD decoder (list size = 2), \cite{list_alex} respectively.

\begin{figure}[t]
\includegraphics[width=1\columnwidth]{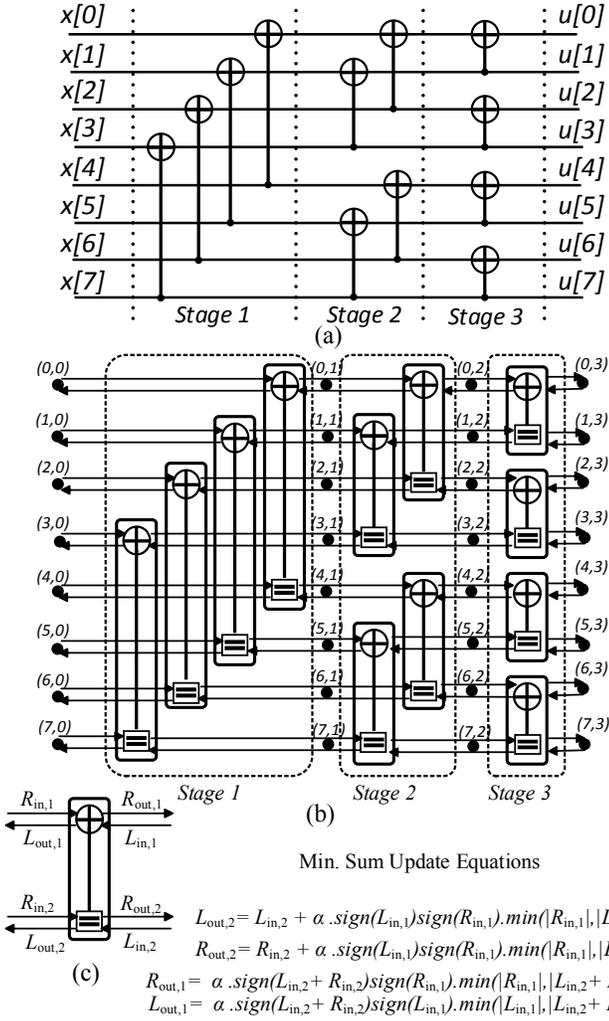}

\caption{\label{fig:Encoding-signal-flow}(a) Encoding signal flow graph of
(8,4) polar code (b) Factor graph of (8,4) polar code. (c) Processing
Element for BPD and min-sum update equations}
\end{figure}

\section{\label{sec:PRELIMINARIES}PRELIMINARIES}

\subsection{\label{subsec:Polar-Codes-and}Polar Codes and Belief Propagation
Decoding}

Polar codes are specified by a generator matrix $\mathbf{G_{n}=F^{\otimes m}}$,
where $n=2^{m}$ is the code length and $\mathbf{F}^{\otimes m}$
is the $m^{th}$ Kronecker power of $\mathbf{F}=\left[\begin{array}{cc}
1 & 0\\
1 & 1
\end{array}\right]$. An $(n,k)$ polar code can be generated in two steps. First, an
$n$-bit message $\mathbf{u}$ is constructed by assigning the $k$
reliable and $(n-k)$ unreliable positions as information bits and
frozen bits, respectively. The $(n-k)$ frozen bits are forced to
0 and form the frozen set $\mathcal{A^{C}}$. Then, the $n$-bit $\mathbf{u}$
is multiplied with the generator matrix $\mathbf{G=F^{\otimes m}}$
to generate an $n$-bit transmitted code-word $\mathbf{x}$. Fig.
\ref{fig:Encoding-signal-flow}(a) shows the encoding signal flow
graph for $n=8$ polar codes, where the \textquotedblleft $\oplus$\textquotedblright{}
sign represents the XOR operation. 

Due to the polarization phenomenon of polar codes, the bit channels
($u_{i},\;i\epsilon\{1,2,...,n\}$) either become completely noiseless
(termed as \textquotedblleft \emph{good channels}\textquotedblright{}
for future reference) or completely noisy (termed as \textquotedblleft \emph{bad
channels}\textquotedblright{} ). Bit channel qualities are measured
by the corresponding Bhattacharyya parameter $Z(u_{i}),\;i\epsilon[n]$,
where $Z(u_{i})$ corresponds to Bhattacharyya parameter of the channel
seen by the bit $u_{i}$ (suitably defined in \cite{arikan}). Lower
valuse of $Z$ mean the corresponding bit channels have very small
error probability, and hence they are known as good channels, and
are used to carry information bits. On the other hand, higher values
of $Z$ imply that the corresponding bit channels have higher error
probability, and thus they are bad channels, and are used for frozen
bits.

Polar codes can be decoded by applying a BP algorithm over the corresponding
factor graph \cite{Arikan-BP}. Similar to the encoding signal flow
graph, the factor graph for an $(n,k)$ polar code $(n=2^{m})$, is
an $m$-stage network which consists of $n\times(m+1)$ nodes. Each
node in the factor graph is associated with a right-to-left and a
left-to-right likelihood message. $L_{i,j}^{t}$ and $R_{i,j}^{t}$
denote the right-to-left and left-to-right likelihood messages of
the $i^{th}$ node at the $j^{th}$ stage and the $t^{th}$ iteration,
respectively. Fig.\ref{fig:Encoding-signal-flow} (b) shows an example
of a three-stage factor graph for $n=8$ polar codes. During the BP
decoding procedure, these messages are propagated and updated among
adjacent nodes using the min-sum updating rule \cite{G-matrix}, as
shown in Fig.\ref{fig:Encoding-signal-flow} (c). 
\begin{figure}[t]
\includegraphics[width=1\columnwidth]{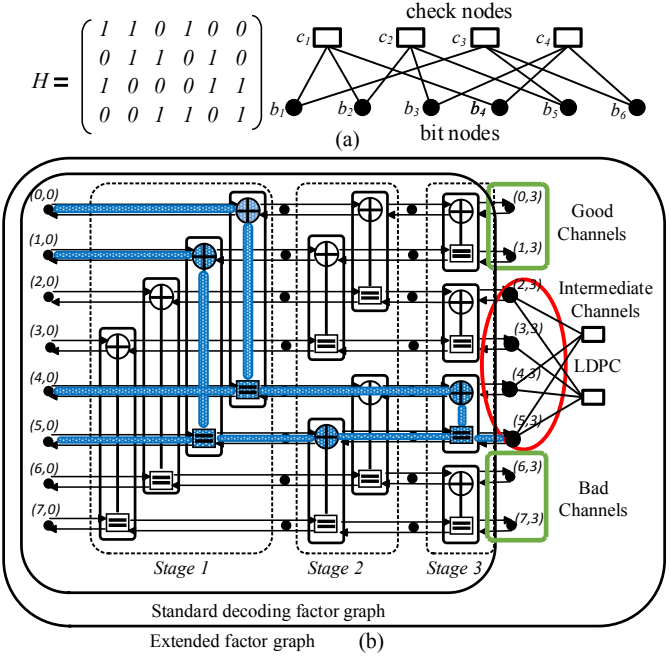}

\caption{(a) Parity check matrix ($H$), for LDPC code of length 6 and its
tanner graph representation\cite{ldpc-sarah} (b) Extended factor
graph for concatenated polar codes (IC-LDPC Polar codes \cite{Guo}
); The stopping tree for node $(5,3)$ is shown in blue, with leafset
nodes \{$(0,0),(1,0),(4,0),(5,0)$\} \label{Fig: Extended-factor-graph}
; Leafset size for $(5,3)$ = 4, equal to row-weight of $6^{th}$
row of $G_{8}$}
\end{figure}

\subsection{\label{subsec:LDPC-codes}Low-Density Parity Check Codes: }

LDPC codes \cite{ldpc} are block codes characterized by a parity
check matrix ($H$), with a constraint on a codeword $(\boldsymbol{x})$
such that $H\boldsymbol{x}^{t}=0$. The $H$ matrix of an example
LDPC of code length 6 is shown in Fig \ref{Fig: Extended-factor-graph}
(a). LDPC codes are often represented in graphical form by \emph{Tanner
graph,} where rows of $H$ correspond to the \emph{check nodes} and
columns of $H$ correspond to the \emph{bit nodes,} respectively.
An edge connects a check node $i$ with a bit node $j$, if and only
if $H_{ij}$ = 1. The number of edges ($e$) present in the\emph{
tanner graph} are equal to the number of 1's present in the parity
check matrix. For the example shown in Fig \ref{Fig: Extended-factor-graph}
(a), $e=12.$ LDPC codes are usually decoded by applying iterative
belief propagation algorithm on their tanner graph, where soft messages
are propagated between the bit nodes and the check nodes in an iterative
manner.

\subsection{\label{IC-LDPC}Intermediate Channel LDPC Polar codes \cite{Guo}:}

Apart from the good or bad channels, there are a smaller number of
bit channels which are not either completely noise free or completely
noisy, hence they are called \textquotedblleft \emph{Intermediate
channels}\textquotedblright . For a given $(n,k)$ polar code with
information set $\mathcal{A},$ intermediate channels correspond to
those information bits which have relatively larger values of $Z(u_{i}),\;i\epsilon\mathcal{A}$
among all information bits. Guo et al. \cite{Guo}, proposed to apply
a shorter outer LDPC code on these intermediate channels, to provide
extra protection on these specific channels so that the overall performance
of Polar codes can be improved. We call this approach as \emph{Intermediate
Channel LDPC Polar codes (IC-LDPC Polar codes)} for future reference. 

Let $ng$ denote the number of good channels and $\nabla n$ denote
the number of channels on which outer LDPC code is applied (these
channels are termed as $u_{ldpc}$ for future reference), then the
rate of polar code is calculated as: $R_{polar}=\frac{ng+\nabla n}{n}$
and the size of the information set $\mathcal{A}$ is: $|A|=ng+\nabla n.$
The rate of overall concatenated LDPC-Polar code will be: $R=\frac{ng+(\nabla n\times R_{ldpc})}{n}$,
where $R_{ldpc}$ is the rate of the outer LDPC code. Since both polar
code and LDPC code can be decoded by belief propagation algorithm
so the factor graph of a polar code can easily be extended to include
the tanner graph of shorter LDPC codes for decoding as shown in Fig.
\ref{Fig: Extended-factor-graph} (b), where $n=8$, $ng=2$, $\nabla n=4$,
and $R_{ldpc}=0.5$ respectively.

\section{\label{sec:Proposed-Concatenated-LDPC-Polar}Proposed Concatenated
LDPC-Polar code}

We propose to use a different criterion to choose the set of bits
to be protected by the outer LDPC code $u_{ldpc}$, based on the notion
of smaller leafset size. 

\subsection{Leafset Size for Information Bits \cite{Eslami}:}

Eslami et al. \cite{Eslami}, analyzed stopping trees as well as girth
of polar codes and their effects on the performance of BPD. Every
information bit in $\mathcal{A},$ has a unique stopping tree rooted
at that information bit (the right hand side of the factor graph)
with its leaves at the code-bits (the left hand side of the graph).
Fig. \ref{Fig: Extended-factor-graph}(b) shows a stopping tree rooted
at an information bit $(node\;(5,3))$ as well as the corresponding
leaf set ($nodes\;(0,0),(1,0),(4,0),(5,0)$ ). Hence every information
bit has an associated leaf set, and the number of code-bits in that
leafset is called leafset size for that information bit. The leaf
set size of the information bit $(node\;(5,3))$, shown in Fig. \ref{Fig: Extended-factor-graph}(b),
is 4. For two information bits with different leafset sizes, under
belief propagation decoding, the one with smaller leafset size is
more likely to be erased than the information bit with larger leafset
\cite{Eslami}.

\subsection{\label{subsec:Minimum-Leaf-LDPC}Proposed Criterion for Choosing
Bits for Outer LDPC Codes }

Due to the significance of the information bits with smaller leafset
size, we propose to choose bits with smaller leafset sizes to be protected
by outer LDPC code ($u_{ldpc}$). To simplify the calculation of leafset
size for each information bit $u_{i},\;i\epsilon\mathcal{A}$ , we
exploit the property that the leafset size of the $i^{th}$ information
bit is equal to the weight of the $i^{th}$ row of generator matrix
$\boldsymbol{G_{n}}$ (Fig. \ref{Fig: Extended-factor-graph}(b)).
Hence we will use leafset size and weight of information bit $u_{i},\;i\epsilon\mathcal{A},$
interchangeably, for the rest of the discussion.\textbf{ }The pseudocode
for choosing $u_{ldpc}$ is presented in Algorithm \ref{algorithm:pseudocode}. 

\begin{algorithm}     
\SetKwInOut{Input}{Input}     
\SetKwInOut{Output}{Output}  
\Input{$\mathcal{A}$, $(Z(u_i), \;i\;\epsilon\;\mathcal{A})$, $\nabla$$n$ and $\mathbf{G_n}$ }     
\Output{$u_{ldpc}$}   
Divide $\mathcal{A}$ into different subsets $\mathcal{A}_i,\mathcal{A}_j$ ... such that each subset contains bits with same row-weight $w$, and $w_i$ <  $w_j$ if $i < j$\;
Sort bits in each subset $\mathcal{A}_k$ in descending order based on Bhattacharyya parameter $(Z(u_i), \;i\;\epsilon\;\mathcal{A}_k\;and\;k=1,2.. )$ such that first bit in each subset is the least reliable bit among all other bits in that subset\;
Choose first $\nabla$$n$ bits from the set \{$\mathcal{A}_i,\mathcal{A}_j$ ...\}\;
\caption{Choosing bit channels for outer LDPC code ($u_{ldpc}$)} 
\label{algorithm:pseudocode}
\end{algorithm}

For (1024,544) polar codes with $ng=480$ and $\nabla n=64$ polar
code rate can be calculated as $R_{polar}=\frac{480+64}{1024}=0.53125$,
with $|\mathcal{A}|=480+64=544$. These 544 information bits contain
bits with weight 16, 32, 64, 128, 256, 512 and 1024, hence the minimum
weight for $\mathcal{A}$ is 16. For choosing $u_{ldpc}$ , these
544 information bits are divided into subsets $\mathcal{A}_{1},\mathcal{A}_{2}...\mathcal{A}_{7}$
such that $\mathcal{A}_{1}$ contains all information bits with minimum
weight 16, similarly $\mathcal{A}_{2}$ contains all information bits
with second minimum weight 32 and so on. The size of $\mathcal{A}_{1}$
is 31 bits whereas the size of $\mathcal{A}_{2}$ is 144. Then each
of these subsets $\mathcal{A}_{1},\mathcal{A}_{2}...\mathcal{A}_{7}$
are sorted in descending order based on Bhattacharyya parameter( $Z(u_{i}),\;i\epsilon[\mathcal{A}_{k}]$$\;and\;k=1,2...7$).
Thus the first bit in each subset has the highest value of $Z$ and
hence corresponds to the least reliable bit among all other bits in
that subset $\mathcal{A}_{k}$$\;where\;k=1,2...7$. Finally first
64 bits are chosen from the set $\{\mathcal{A}_{1},\mathcal{A}_{2}...\mathcal{A}_{7}\}$,
such all 31 bits of $\mathcal{A}_{1}$are chosen and first 33 bits
of $\mathcal{A}_{2}$, which are the least reliable bits of $\mathcal{A}_{2}$,
are selected.

\subsection{Scheduling Scheme for Concatenated LDPC-Polar Code:}

For the proposed concatenated LDPC-Polar code, round-trip scheduling
\cite{round-trip} is employed. For the extended factor graph shown
in Fig. \ref{Fig: Extended-factor-graph} (b), one iteration in round-trip
scheduling is completed when the information from the left side of
the factor graph \emph{(i.e. the channel LLR)} travels all the way
to the right side of factor graph where it is passed as \emph{intrinsic
(a priori) }information to the the tanner graph of the LDPC, where
\emph{bit nodes} to\emph{ check nodes} messages are calculated and
propagated toward the \emph{check nodes}. Following this leftward
information flow, \emph{check nodes} to \emph{bits nodes} message
are calculated and passed to \emph{bit nodes}. This \emph{extrinsic}
information from the tanner graph along with the frozen bit information
of polar codes is propagated rightward towards the right side of factor
graph and hence one iteration is completed.

\subsection{Complexity Analysis}

In comparison with original BPD scheme, i.e., without LDPC concatenation
(termed as baseline BPD), both IC-LDPC Polar codes and proposed LDPC-Polar
codes have higher complexity due to the inclusion of tanner graph
with the factor graph as shown in Fig. \ref{Fig: Extended-factor-graph}
(a). Complexity for the baseline BPD, for one iteration in the round-trip
scheduling, is the summation of the complexity for leftward message
propagation and the complexity for rightward message propagation and
is equal to $4\times n\,log\,n$ additions where $n$ is the length
of polar code. For simplicity, one minimum operation is also counted
as one addition. For (1024,512) polar code, the complexity of baseline
BPD per iteration will be 40960 additions.

For the extended tanner graph, the complexity for one iteration will
be the summation of the complexity of the baseline BPD and the complexity
of the tanner graph. The complexity for the tanner graph can be calculated
as the summation of the complexity for the bit nodes to check nodes
message propagation and that for the check nodes to bit nodes message
propagation. For a $regular\;(3,6)$ LDPC code with $lb$ code bits
and $lc$ check bits, the complexity for bit nodes to check nodes
message propagation is $2e$ where, $e$ is the number of edges present
in the tanner graph as mentioned in \ref{subsec:LDPC-codes}. Similarly
the complexity for check nodes to bit nodes message propagation is
also $2e$. For (1024,544) polar code and a regular (3,6) LDPC code
with code bits = 64 ($lb$) and check bits = 32 ($lc$), the number
of edges present in the tanner graph ($e$) is 192 and the complexity
for one iteration in the round-trip scheduling is thus equal to 40960
+ 384 + 384 = 41728 additions. Hence the concatenated LDPC-Polar code
design incurs just a small complexity overhead of 1.84 \% per iteration,
over baseline BPD. 
\begin{figure}[t]
\includegraphics[width=1\columnwidth]{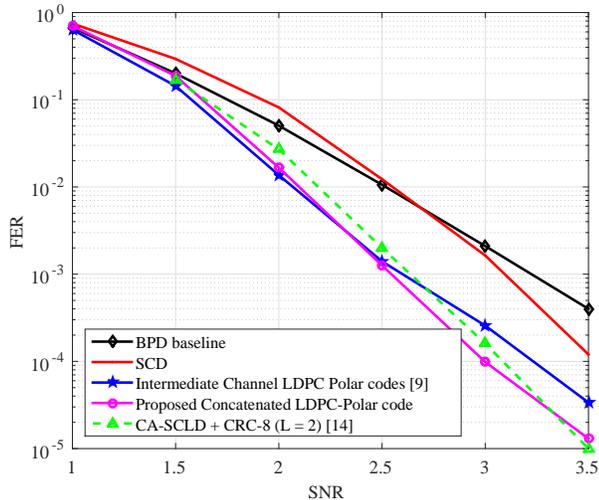}

\caption{\label{fig:simulation-results}Error correcting performance with maximum
number of iterations = 60.}
\end{figure}

\section{\label{sec:SIMULATION-RESULTS}SIMULATION RESULTS}

We carried out simulations on polar code of length $(n)$ \emph{1024}
and rate ($R$) \emph{0.5} and compared the performance of the proposed
concatenated LDPC-Polar code with the baseline BPD and the IC-LDPC
Polar codes. Fig. \ref{fig:simulation-results} shows the simulation
results over an AWGN channel with BPSK modulation. For all three BPD
implementation, scaled min-sum approximation with scaling parameter
( \textgreek{a}=0.9375 ) was used for the update equations Fig.\ref{fig:Encoding-signal-flow}
(c). Similarly for all three BPDs, round-trip scheduling was employed
and the maximum number of iteration are set to 60. 

For both IC-LDPC Polar codes and proposed concatenated LDPC-Polar
code, $ng=480$ , $\nabla n=64,$ and a regular $(3,6)$ LDPC code
with $code\;bits=64$ and $parity\;bits=32$ ($R_{ldpc}=0.5$) is
used as an outer LDPC code such that the overall concatenated LDPC-Polar
rate ($R$) is $0.5$ and $R_{polar}=0.53125$. Fig. \ref{fig:simulation-results}
shows the simulation results. It can be seen that the proposed concatenated
LDPC-Polar code results in 0.25dB and 0.5dB performance improvement
at $10^{-4}$ when compared with IC LPDC Polar code \cite{Guo} and
SCD respectively. Moreover, we have also compared with \emph{CRC-Aided
Successive Cancellation List Decoder (CA-SCLD)} with list size = 2
\cite{list_alex}. It is to be noted that, the overall code rate of
\emph{CA-SCLD} is 0.5 and it employs CRC-8 with $list\;size=2$. The
proposed LDPC-Polar code has a performance improvement of 0.1dB at
$10^{-4}$ over \emph{CA-SCLD. }Moreover with \emph{CA-SCLD} CRC-8
takes a latency of 2660 cycles to decode one frame \cite{list_alex}
whereas, due to highly parallel nature of BPD, the proposed LDPC-Polar
codes will result in much lower latency. 

\section{\label{sec:CONCLUSION}CONCLUSION }

In this work, we have presented a novel concatenated LDPC-Polar code,
where a small outer LDPC code is concatenated with a larger inner
polar code. Information bit channels with smaller leafset size are
proposed to be protected by outer LDPC code. The proposed concatenated
LDPC-Polar code results in 0.5dB, 0.25dB and 0.1dB performance improvement
at $10^{-4}$ over SCD, an existing concatenated LDPC-Polar code approach
and the state-of-the-art list decoder, respectively. Moreover the
proposed concatenated LDPC-polar code only incurs a small complexity
overhead of 1.84\% per iteration, compared to baseline BPD. For future
works, we intend to apply early stopping methods to further reduce
the latency of decoding, hence to increase the throughput.


\begin{thebibliography}{10}
\bibitem{arikan}E. Arikan, \textquotedblleft Channel polarization:
A method for constructing capacity achieving codes for symmetric binary-input
memory less channels,\textquotedblright{} \emph{IEEE Trans. Inf. Theory},
vol. 55, no. 7, pp. 3051-3073, 2009.

\bibitem{arikan2} E. Sasoglu, E. Telatar, and E. Arikan, \textquotedblleft Polarization
for arbitrary discrete memoryless channels,\textquotedblright{} in
\emph{Proc. IEEE Inf. Theory Workshop (ITW)}, 2009, pp. 144\textendash 148.

\bibitem{yazdi}A. Alamdar-Yazdi and F. R. Kschischang, \textquotedblleft A
simplified successive cancellation decoder for polar codes,\textquotedblright{}
\emph{IEEE Commun. Lett.}, vol. 15, no. 12, pp. 1378\textendash 1380,
Dec. 2011.

\bibitem{scdarch}C. Leroux, I. Tal, A. Vardy, and W. J. Gross, \textquotedblleft Hardware
architectures for successive cancellation decoding of polar codes,\textquotedblright{}
\emph{in Proc. IEEE Int. Conf. Acoust., Speech, Signal Process. (ICASSP)},
May 2011, pp. 1665\textendash 1668,.

\bibitem{Arikan-BP}E.Ar\i kan, \textquotedblleft A performance comparison
of polar codes and Reed-Muller codes,\textquotedblright{} \emph{IEEE
Commun. Lett.}, vol. 12, no. 6, pp. 447\textendash 449, Jun. 2008. 

\bibitem{TVLSI} S. M. Abbas, Y. Fan, J. Chen and C. Y. Tsui, \textquotedbl{}High-Throughput
and Energy-Efficient Belief Propagation Polar Code Decoder,\textquotedbl{}
\emph{in IEEE Transactions on Very Large Scale Integration (VLSI)
Systems, vol. 25, no. 3, pp. 1098-1111, March 2017.}

\bibitem{Jun_lin} J. Lin, C. Xiong and Z. Yan, \textquotedbl{}Reduced
complexity belief propagation decoders for polar codes,\textquotedbl{}
\emph{in Proc. IEEE Workshop on Signal Processing Systems (SiPS),
1-6 October 2015}.

\bibitem{Eslami} A. Eslami and H. Pishro-Nik, \textquotedbl{}On finite-length
performance of polar codes: Stopping sets, error floor, and concatenated
design,\textquotedbl{} \emph{in IEEE Transactions on Communications,
vol. 61, no. 3,, March 2013, pp. 919-929.}

\bibitem{Guo}J. Guo, M. Qin, A. Guillén i Fàbregas and P. H. Siegel,
\textquotedbl{}Enhanced belief propagation decoding of polar codes
through concatenation,\textquotedbl{} \emph{in Proc. IEEE International
Symposium on Information Theory 2014 , Honolulu, HI, 2014, pp. 2987-2991.}

\bibitem{G-matrix}B. Yuan and K.K. Parhi, \textquotedbl{}Early stopping
criteria for energy-efficient low-latency belief-propagation polar
code decoders,\textquotedbl{} \emph{IEEE Trans. Signal Process.},
vol.62, no.24, pp.6496\textendash 6506, Dec.15, 2014. 

\bibitem{round-trip}J. Xu, T. Che and G. Choi, \textquotedbl{}XJ-BP:
Express journey belief propagation decoding for polar codes,\textquotedbl{}
\emph{in Proc. IEEE Global Communications Conference (GLOBECOM) 2015,
San Diego, CA, 2015, pp. 1-6.}

\bibitem{ldpc} R. G. Gallager, \emph{Low Density Parity-Check Codes}.
MIT Press, Cambridge, MA, 1963.

\bibitem{ldpc-sarah} Sarah J. Johnson and Steven R. Weller, \emph{Low-Density
Parity-Check Codes: Design and Decoding}, Wiley Encyclopedia of Telecommunications,
2003, pp.1-18.

\bibitem{list_alex}A. Balatsoukas-Stimming, M. B. Parizi and A. Burg,
\textquotedbl{}LLR-based successive cancellation list decoding ofpolar
codes,\textquotedbl{} \emph{IEEE Transactions on Signal Processing},
vol. 63, no. 19, pp. 5165-5179, Oct.1, 2015.
\end{thebibliography}
\end{document}